# Projection and Invariance in Scientific Explanation

**Harry Sticker**
**Ganymede Technology**
**New York, New York, USA**
**hsticker@ganymedetechnology.com**

## Abstract

Any representational enterprise must omit variation in order to function. NASA still uses Newtonian mechanics, though Einstein superseded Newton — and the standard picture of scientific progress cannot explain how. A description that omitted nothing would be identical to its subject and would explain nothing. This paper argues that omission is not a defect but the central structural feature of any enterprise that builds representations from incomplete information. The key concept is *projection*: a principled mapping from underlying complexity to a descriptive space that partitions states into equivalence classes, omits within-class variation, and makes patterns visible that would otherwise be lost. Projection is simultaneously revelatory and constitutive: it makes genuine invariants tractably accessible while bringing into being the concepts through which they become expressible. The paper distinguishes vertical cases, in which earlier projections survive as limiting cases of more refined successors with recoverable omission, from horizontal cases, in which omission is constitutive and invariants are accessible only at the level of the projection that defines them. The framework accounts for persistent pluralism in mature sciences, treats the renormalization group as a systematic implementation of the invariant-tracking criterion, and defends a level-relative realism on which higher-level projections reveal genuine structural features of the world. The deepest claim is an inversion of the standard picture: perspectival structure is not a concession to complexity but the condition for invariant detection. A world rich in invariants cannot be exhausted by a single projection.

## 1. The Puzzles

Before you flip a coin, the probability of heads is one-half. After you flip it, the probability is either zero or one. The coin lands heads, and the outcome was, in principle, determined before you let go — the trajectory, the angular momentum, the surface, all fixed by prior physical conditions. So what was that one-half? It was real enough to build a casino on. And then it vanished the moment the coin hit the table. Where did it go?

The standard answers are unsatisfying. If the probability was a feature of the world, it disappeared the moment the coin landed — strange behavior for a feature of the world. If it were only a feature of our ignorance, then the profits of casinos are built on ignorance rather than on fact. Both answers treat probability as a binary choice: fully real or merely subjective. But the coin flip suggests a third possibility — one that neither answer can reach.

In 1687, Newton published a theory of motion and gravitation that unified the trajectories of planets, projectile motion, and tidal behavior within a single mathematical framework. In 1915, Einstein showed that Newtonian mechanics was incomplete. General relativity replaced it with a superior account of gravity, space, and time. And yet NASA uses Newtonian mechanics to plot spacecraft trajectories. The calculations are accurate. The landings succeed. If Newton's theory was superseded — if it was wrong — why does it keep working?

Such persistence is not an isolated curiosity. It is the normal condition of theoretical knowledge. Lavoisier overturned the phlogiston theory, yet phlogiston heat-flow mathematics continued to predict experimental results long after the caloric fluid was abandoned. Ptolemaic astronomy was displaced by Copernican heliocentrism, yet Ptolemaic methods remained accurate for practical navigation for generations. Something is being preserved across these transitions that the standard picture of progressive replacement cannot explain. The standard picture says that the superseded theories are simply wrong. But if they are simply wrong, why do they keep working?

A second puzzle that compounds the first. If science converges on single correct descriptions of its domains, mature sciences should exhibit increasing theoretical unification. What we observe is almost the opposite. In biology, over twenty competing definitions of species remain in active scientific use (Mayden 1997). In personality psychology, the Big Five factor model and the HEXACO framework each divide the space of personality variation differently, each productive, neither definitively correct (John et al. 2008). These coexisting frameworks are not symptoms of immaturity. They are marks of fields doing serious scientific work. Practitioners do not experience pluralism as a failure. Biologists do not believe taxonomy is in crisis. The frameworks coexist productively, each illuminating something the others do not. Something other than convergence is going on — and the standard picture has no account of it.

A third phenomenon points to the same underlying structure, though from an unexpected direction. In 1668, Edme Mariotte discovered that the human retina has a region with no photoreceptors — the point where the optic nerve exits — and that this gap in the visual field is entirely invisible to the perceiver (Mariotte 1668). The visual system does not flag the missing region as unknown. It fills it in seamlessly. What is left out does not appear as an absence. It simply does not appear at all.

The visual blind spot is not a puzzle about vision. It is a structural fact about representational systems: a system can be constitutively unable to register its own omissions as omissions. What is left out does not appear as a gap. It appears as a continuation — or not at all. This suggests that the invisibility of limits is not a contingent failure of particular theories but a structural feature of representation itself. If so, it belongs with the other two puzzles, not beside them.

Three questions then call for a common answer: Why do superseded frameworks continue to work within their original domains? Why do mature sciences sustain multiple incompatible frameworks without this constituting a failure? And why do representational systems so often fail to register the limits of their own access? These questions look different. They are not. Behind all three lies the same structural feature — one that the standard picture of progressive replacement cannot see, for the same reason that any representational system cannot fully see its own omissions. This paper names that feature and shows what follows from it.

## 2. Projection

In order to function, every representation must omit variation. These omissions are not a limitation to be overcome. They are the condition of representation itself. A description that omitted nothing would be identical to its subject — not a description of the phenomenon but a reproduction of it. Jorge Luis Borges imagined a map drawn at a scale of one mile to the mile: perfectly accurate, perfectly complete, and perfectly useless (Borges 1946). The tattered remnants were eventually abandoned in the desert. A map that omits nothing is not a better map. It is not a map at all.

What the three puzzles described above share is a specific kind of omission — one that is principled rather than accidental, and structural rather than incidental. A theory does not omit variation randomly. It groups underlying states that are equivalent for its purposes, omits variation within those groups, and in doing so makes certain patterns visible that would otherwise be lost in detail. This operation — partitioning, omission, and the making-visible of invariants — is what this paper calls projection.

A projection is a principled mapping from underlying complexity to a structured descriptive space that (1) partitions underlying states into equivalence classes, (2) omits variation within those classes, and (3) thereby makes certain patterns visible that would otherwise be lost. These three components constitute the definition, and each does real work.

***Partitioning***. A projection groups underlying states that are treated as equivalent for some purpose. In the visual system, many different surface reflectances under varying illumination are grouped as the same color — variation in raw luminance is irrelevant to the stable surface property being tracked. In population genetics, many different molecular variants are grouped into the same fitness class. In economics, many different transaction configurations are aggregated into a single quantity. In each case, the grouping is constrained by the invariant structure of the domain rather than freely imposed. In physical theory, this operation is called coarse-graining — the partitioning of a fine-grained state space into coarser equivalence classes — and it is the standard physical instance of what projection does.

***Omission***. Variation within an equivalence class is absent from the representation — not flagged as absent, simply absent. The projection carries no internal marker of what it has left out. This is the source of seamlessness: from within the projection, the omitted variation does not register as missing. Mariotte's blind spot does not appear as a filled gap. It does not appear at all. A physicist working within Newtonian mechanics does not experience the limits of the framework as limits. The limits do not arise as representable features of the domain.

***Availability.*** The projection makes a pattern accessible in a form usable for the relevant explanatory task — one that would not be available without the projection's organizing work. Temperature is not available for thermodynamic reasoning without a projection that groups molecular configurations by mean kinetic energy. Fitness is not available for evolutionary reasoning without a projection that groups organisms by heritable traits and differential reproductive success. Availability names what the projection is for: making structure tractably accessible that the underlying complexity alone cannot supply. Availability is not merely computational convenience. It is the condition under which the invariant can enter into stable explanations, predictions, and interventions.

The definition uses a term that requires explanation: *invariant*. It could be read as naming whatever a projection happens to preserve — in which case any projection would trivially reveal its own invariants, and the framework would collapse into relativism. That is not the intended reading. An invariant is not an output of a projection. It is what resists variation across projections. A genuine invariant persists when the projection changes. An artifact of a projection, by contrast, disappears when the projection changes. It is a feature of the specific description, not of the domain.

***Stability under independently motivated transformations.*** If a regularity holds only under transformations its description was tailored to admit, it is not an invariant. It is a feature of the fitting procedure. A genuine invariant survives transformations that were not chosen with it in mind: reparameterizations grounded in the physics of the domain, coordinate changes motivated by symmetry, coarse-grainings justified by scale separation. What survives is a feature of how the domain responds to perturbation, not a feature of how flexibly the description can be reshaped. In the social and aggregate-physical domains, the relevant transformations are substitutions: changes to the physical substrate that preserve the functional or causal role the projection treats as invariant.

***Cross-model robustness under structurally distinct projections.*** If a regularity is tracked by only one projection of the domain, it may be an artifact of that projection's particular partitioning. A genuine invariant is detectable from structurally different projective angles: a thermodynamic quantity that shows up in statistical mechanics, in hydrodynamics, and in the renormalization group analysis of critical behavior; a selection effect that is visible under Fisherian, coalescent, and genealogical frameworks. The invariant is what the different projections converge on rather than what any one of them uniquely produces.

***Counterfactual support under intervention.*** An invariant supports predictions about what would happen under perturbations that were not part of the evidence used to construct the projection. A curve-fit does not pass this test: by construction, it reproduces the data it was fit to, and its extrapolations are warranted only within the envelope of that data. A genuine invariant licenses counterfactuals that reach outside that envelope — predictions about systems not yet measured, conditions not yet realized, parameter ranges not yet probed. This is a necessary condition for the invariant to enter stable intervention, a condition named Availability above.

Temperature passes all three. It is stable under reparameterizations of the microscopic description; it is robust across models, showing up in statistical mechanics, in thermodynamics, in hydrodynamic limits, and in the renormalization group analysis of phase transitions; and it supports counterfactual predictions about systems and conditions well outside the experimental regimes in which thermodynamic laws were first established. The universality class invariants of Section 4.3 pass even more sharply: they are defined by what survives an unbounded sequence of coarse-grainings — a stringent form of stability under transformation.

By contrast, a numerological system can be tuned to fit any historical dataset — stock prices, planetary periods, crop yields — by adjusting its parameters. Such a system can satisfy empirical adequacy within the domain of its fit, but it routinely fails these diagnostics: its regularities collapse under transformations not anticipated by the fitting procedure, they do not recur when the same domain is projected through an independent framework, and they do not support counterfactual predictions outside the envelope of the fit. A more contemporary example: a high-capacity neural network fit to a benchmark dataset can achieve very high in-sample accuracy while failing under distribution shift, transfer to adjacent tasks, or independent retraining on data drawn from the same phenomenon under different collection conditions. The network tracks the dataset. It does not track what the dataset is of.

Failure is not uniform but structured. Different defective descriptions fail different components of the constraint. A simple curve fit to observed data typically satisfies empirical adequacy within its fitting domain but fails counterfactual support: its extrapolations degrade rapidly outside the data envelope, revealing that it has captured correlations without tracking the structure that generates them. Overparameterized machine learning models often exhibit a different failure mode: they may generalize within a narrow distribution but fail to be stable under independently motivated transformations, such as distribution shift, re-sampling, or retraining on equivalent data drawn under different collection conditions. Their apparent success is local to the training regime rather than stable across admissible perturbations. Numerological systems fail more globally: they do not survive independent reparameterization, do not converge across distinct modeling frameworks, and do not support reliable intervention.

These are not variations of a single deficiency but distinct ways of failing to track invariants. The diagnostic, therefore, does not merely separate successful from unsuccessful descriptions. It differentiates modes of failure according to which structural constraint is violated. What survives all three constraints is not simply well-fit, but stable across the independent axes along which artifacts fail. Empirical adequacy is a weaker filter than the diagnostic. Many adequate descriptions do not survive it. What survives counts as tracking a feature of the domain rather than a feature of its description.

The three conditions are jointly necessary and jointly sufficient: what it is for a description to track genuine invariant structure just is for it to survive this filtering regime. The perceptual case of Section 2.1 supports this: a detection system earns the realist description of what it tracks not by correspondence to an external standard but by surviving adversarial filtering of the relevant kind. The scientific case is analogous.

This is why multiple projections of the same domain can reveal different genuine invariants without either being wrong. The domain has a more stable structure than any single projection can make visible. Different projections make different aspects of that structure tractable. The tracking of genuine invariants — not the uniqueness of the projection — is what confers legitimacy. Projection does not merely redescribe a domain; it is constrained by the requirement that the invariants it reveals remain stable under independent perturbation and reparameterization. This constraint is what distinguishes projection from arbitrary redescription.

Omission is constitutive, not incidental. The utility of any representation comes from what it leaves out. A complete microphysical description does not, by itself, make thermodynamic structure available. A complete record of all individual transactions does not contain macroeconomics. A complete trace of every organism does not contain ecology. The level-specific regularities become visible only under projections that omit the detail that would obscure them. Omission is the mechanism of explanation, not its cost.

## 2.1 The Dual Character of Projection

Projection does two things simultaneously, and the two are more intimately connected than they first appear.

The first is *revelatory*. A projection makes genuine invariants tractably accessible — patterns that are real features of the world, not artifacts of the organizing work. Temperature tracks a fundamental property of molecular kinetic behavior. Fitness tracks a real relationship between heritable traits and reproductive outcomes. Universality classes capture a real feature of the behavior of physical systems near critical transitions. Projection does not create these patterns. It selects the equivalence classes under which they become visible.

The second is *constitutive*. The concepts through which invariants become accessible do not pre-exist the projections that define them. Before a projection groups molecular configurations by mean kinetic energy, there is no concept of temperature — there is only molecular motion. Before an economic projection groups physically heterogeneous objects by their functional role as media of exchange, there is no concept of money — there are only metals and paper. The concept arises through the act of defining the equivalence class. To partition is to constitute.

The dual character might seem to create tension. If projections constitute concepts, how can they simultaneously reveal features of the world rather than merely constructing them? The resolution is that the constitution is constrained. A projection is not arbitrary: its equivalence classes are determined by genuine structural stability in the domain, detectable through the diagnostic of the previous subsection. The invariants are discovered. The concepts are constructed. The legitimacy of the concepts depends entirely on whether the construction has tracked what the world keeps stable under independent perturbation.

The epistemological stance this framework presupposes has an uncontroversial precedent. A visual system tracks real surface properties of the distal environment — reflectance, orientation, texture, and relative depth. No one seriously doubts this. But the visual system has no access to those properties independent of its own detection; there is no template against which its outputs can be checked. What warrants the realist description of what it tracks is not correspondence to an external standard, and not predictive success on its own. It is survival under a specific filtering regime: variation in illumination, viewing angle, occlusion, and context. Detectors that did not track stable distal properties would not continue to produce the systematically successful action that selected for them. The warrant is structural. A detection mechanism earns the realist description of what it tracks by surviving adversarial filtering of a specified kind, not by succeeding in general. The detected features are, in that sense, models of the real — structured representations whose claim on reality is certified by surviving adversarial filtering, not by comparison to a ground truth.

Helmholtz recognized this in the nineteenth century, and Marr formalized it in the twentieth (Helmholtz 1867; Marr 1982). Perception is not a passive reception of pre-formed features. It is active construction, constrained by the structure of the distal environment. The perceptual system builds descriptions that make stable properties of the world tractably accessible, and the concepts in which those descriptions are framed are constituted by the constructive operations that produce them. Vision is thus the existence proof that a detection system can warrant claims about real features without independent access to those features — and that the warrant, once earned, transfers to any detection system operating under analogous filtering.

The projection framework takes the same epistemological stance, applied to scientific invariants rather than perceptual features. The filtering regime is different in content. Where the perceptual system faces variation in illumination and viewing angle, the scientific projection faces independent reparameterization, comparison across structurally distinct models, counterfactual probing under intervention, and the slow accumulation of anomalies that no internal adjustment can accommodate. Projections that survive that regime earn the same structural warrant that perceptual features earn under selection. Whether the two cases share biological machinery is a separate question the framework does not need to answer.

What matters is that the stance is not invented for the scientific case. It is already in place for a case no one disputes. The diagnostic of the previous subsection articulates, for scientific projections, filtering conditions structurally analogous to those under which perceptual features earn their realist description.

Empirical adequacy is not enough. A description can fit its data perfectly and still fail the diagnostic — its regularities can collapse under transformations the fitter did not anticipate, vanish when the same domain is projected through an independent framework, and break down outside the envelope of the fit. What survives all three conditions is doing something the description alone cannot do: it is tracking what the domain keeps stable across representational variation. That is the step beyond adequacy, and it is taken not by adding metaphysical commitment but by noticing what kind of property cross-model convergence is. It is not a property that the description can generate on its own. It is one that the domain must supply.

A consequence follows that resolves the third puzzle from Section 1. Because a projection omits variation without carrying any internal record of that omission, the omission is invisible from within. The physicist reasoning about temperature does not experience it as a projection that groups molecular configurations — she experiences temperature as a direct feature of the world. The categories of any active projection feel like the natural joints of reality rather than constrained choices about what to omit. The underlying states compatible with a given higher-level description but differing in every omitted detail do not appear as a hidden space of alternatives. They do not appear at all.

The omissions of any projection are invisible from within. The visual system does not flag the missing region as unknown. It fills it seamlessly. A scientific framework does not flag its limits as limits. They appear as natural boundaries — as the way things simply are. The limits of a projection are visible only from outside, typically from a different projection that reaches where the first cannot, or from the accumulation of anomalies that no internal adjustment can resolve.

The Mariotte puzzle from Section 1 resolves here. The visual blind spot is not an isolated curiosity about retinal anatomy. It is a structural instance of a general feature of representational systems: constitutive omission is invisible to the system that performs it. Scientific frameworks are subject to the same condition. That is not a defect. It is what makes them functional.

### 2.2 The Two-Component Distinction

Every theoretical framework has two separable components that can come apart in both logical and historical terms.

The first is the *representational structure*: the variables chosen, the equivalence classes defined, and the invariants those classes make visible. This structure is what makes patterns tractably accessible for inquiry. When a framework successfully tracks genuine invariants, its representational structure tends to survive theory change.

The second is the *substrate interpretation*: the account of what underlies the projection — what the variables refer to at a deeper level, what mechanisms produce the invariants the framework reveals. This component is more vulnerable. It may be replaced entirely while the representational structure survives intact.

Galen's four-temperament structure survived two millennia of medical practice despite the abandonment of the humoral physiology supposed to explain it. The caloric mathematics of heat flow continued to generate accurate predictions long after the caloric fluid was discarded. In each case, the representational structure outlasted the substrate interpretation because it tracked genuine invariants — invariants that passed the diagnostic under successive reformulations — regardless of whether the underlying story was correct. A projection can be explanatorily real before its causal basis is understood.

### 2.3 Further Principles and Adjacent Accounts

A second consequence of the definition: because a projection maps many underlying states onto fewer representational variables, the mapping underdetermines which higher-level organization is legitimate. Many incompatible projections can fit the same evidence. This is the Duhem-Quine point generalized: evidence alone does not fix which projection to adopt. Fitting the available evidence is necessary but not sufficient. Many-to-one mappings underdetermine admissible organization. The consequence is that multiple projections of the same domain can be simultaneously legitimate — not because the domain is too complex for a single correct description, but because different projections make different genuine invariants tractably accessible.

Deduction sits at the boundary of this picture. A valid deductive argument introduces no new equivalence classes: given the premises, exactly one conclusion follows by necessity. Deduction is the limit case of the projective continuum relative to a fixed representational structure — the point at which the contingent space has been reduced to zero by prior projective operations, and nothing remains but transparent derivation. Every inference that goes beyond its premises — induction, abduction, analogy, inference to the best explanation — adds new equivalence classes and is therefore projective. Theory building is irreducibly projective. The view from nowhere is unavailable not as a practical limitation but as a structural one.

The concept of projection relates to several adjacent accounts in order of priority: projection determines the descriptive space within which those operations are performed. *Idealizations* simplify constructs already defined by a projection. *Abstractions* omit detail from a representational space already in place. *Models* explore regions of a space already defined. All three presuppose a projection. They do not constitute one.

*Perspectivalism* holds that science uses multiple perspectives, each capturing something real (Massimi 2022; Giere 2006). This view is correct but descriptive rather than explanatory. It does not say what structural features make a representation perspective-dependent, why earlier perspectives persist within successor ones, or why certain invariants are tied to particular levels of description. The projection framework supplies that structure: a projection does not merely adopt a perspective, it partitions a state space under constraints imposed by genuine invariant structure, and that partition constitutes the invariants it reveals.

Massimi's perspectival realism develops the most systematic recent defense of perspectival knowledge, arguing that perspectival truths are genuinely true and that perspectives can be coordinated across scientific communities without collapsing into relativism (Massimi 2022). The projection framework is not a rival to this account. Its target is different: not the epistemic standing of perspectival claims but the structural operation that generates a perspective in the first place. What Massimi's account does not supply — and what the projection framework provides — is a criterion for distinguishing a perspective that tracks genuine invariants from one that tracks artifacts of the description. The three-condition diagnostic distinguishes these cases. A perspective earns its epistemic standing, on the present account, by satisfying that diagnostic: surviving independently motivated transformations, converging across structurally distinct frameworks, and supporting counterfactuals beyond its fitting domain. This is not a correction of perspectival realism. It is a specification of the structural mechanism that perspectival realism needs but does not itself provide.

*Structural pluralism* holds that scientific pluralism is a permanent and warranted feature of mature inquiry, grounded in the complexity of natural systems or the disunity of natural kinds (Mitchell 2003; Dupré 1993; Kellert, Longino, and Waters 2006). The projection framework accepts this conclusion while offering something additional: a structural account of why legitimate pluralism has the form it does. Multiple projections can be simultaneously warranted not because systems are complex or kinds disunified — though they may be — but because distinct projections define distinct equivalence classes and thereby make distinct invariants visible. Not every pluralism is principled. A projection that fails the invariance diagnostic is not a legitimate alternative framework. It is a defective one.

What projection adds beyond these accounts is a specification of the structural operation — partitioning under constraint — that generates legitimacy and explains its limits.

## 3. Vertical Cases

If a projection tracks genuine invariants, what happens when a better projection arrives? The natural expectation is replacement: the old projection is shown to be wrong, the new one takes its place, and the transition marks progress. This expectation is wrong — or rather, it is right about the substrate interpretation and wrong about the representational structure. When a projection genuinely tracks invariants, it is not replaced by a successor. It is embedded within one.

A structural claim follows: legitimate projections track invariants, and a successor projection that tracks the same genuine invariants must contain its predecessor as a limiting case. If the earlier projection tracked genuine invariants within its domain — and it did — the successor must recover those results as limits. A successor that simply discards what worked is not progress. It is a replacement that loses real knowledge. The embedding relation is not a courtesy extended to superseded theories. It is a structural requirement imposed by the invariants themselves.

Two historical cases show how this operates.

### 3.1 The Astronomy Sequence

The history of astronomy is usually read as a story of progressive revolution — Copernicus overturning Ptolemy, Kepler overturning Copernicus, Newton overturning Kepler. On the present account, it is something more precise: a structured progression in which each projection embeds its predecessor as a limiting case, and in which the deepest move at each step is not mathematical but ontological.

Ptolemy's projection organized planetary observation around two commitments: celestial bodies move in circles, and Earth sits at the center of the cosmos. Both were ontologically mandated — Aristotelian natural philosophy required circular celestial motion. Ellipses were geometrically available to any astronomer of the period. They were not tried because they were ontologically impermissible. The framework tracked genuine invariants of planetary position, even if its partitioning of orbital shapes fell short.

Copernicus changed the reference frame — placing the Sun at the center, the Earth in orbit — but retained circles. The substrate interpretation was transformed; the partitioning of orbital shapes was unchanged. Ptolemy's projection was embedded rather than discarded.

Kepler made the deeper move. He changed the partitioning itself: from the circle — the degenerate conic section, with eccentricity zero — to the full class of conic sections. The circle is a limiting case of the ellipse. Ptolemaic circles are Keplerian ellipses with eccentricity zero. The earlier projection was not refuted. It was recovered as the special case in which eccentricity vanishes.

The same structure repeats at every subsequent transition. Newtonian mechanics is Einsteinian mechanics at velocities well below the speed of light. Geometrical optics is wave optics at wavelengths short relative to aperture size. In each case, the earlier framework continues to work within its domain because it is the limiting case of the correct projection in that domain. NASA uses Newtonian mechanics not despite Einstein but because of him: the embedding relation guarantees that Newtonian results are recoverable wherever relativistic corrections are negligible.

What Copernicus changed was the substrate interpretation — the account of what sits at the center of the cosmos. What Kepler changed was the representational structure — the geometric primitive used to partition orbital shapes. Both were genuine advances, but they operated at different levels, and only the second produced an embedding relation that recovered the information that came before.

The incommensurability thesis — that successive theoretical frameworks are so different that scientists cannot fully understand one another across their transition, and that no rational basis for theory choice between them exists — has been influential since Kuhn (1970) and Feyerabend (1975). The Keplerian case challenges it directly. The new projection was constructed from within the resources of the old one: Kepler was trained in Copernican astronomy, worked with Brahe's data, and responded to anomalies the Copernican framework itself generated. Representational continuity — the embedding of Ptolemaic circles as limiting cases of Keplerian ellipses — is precisely what makes the transition intelligible rather than miraculous. What the incommensurability thesis correctly identifies is discontinuity at the level of substrate interpretation — the world picture does change sharply across major theoretical shifts. What it misses is the representational continuity that underlies the discontinuity and makes rational theory change possible. The sense of revolution is best understood as the phenomenology of ontological space expansion: the discovery that a previously forbidden partitioning was always geometrically available but ontologically impermissible.

## 3.2 Darwin without Genetics

When Darwin published the Origin in 1859, his projection — heritable variation subject to differential reproductive success — identified the right equivalence classes for understanding biological change and revealed genuine invariants about the structure of evolution (Darwin 1859). The framework has been empirically productive since its publication.

Darwin had no account of the underlying mechanism. Mendel's work on inheritance was published in the same decade; the molecular basis of heredity would not be established until the twentieth century (Mendel 1866). Darwin's projection was entirely correct about the structure of the phenomenon and entirely silent about its causal basis. A projection can be explanatorily real before its substrate is understood.

His projection identified the right equivalence classes — populations, variant traits, differential reproduction rates — and the right invariants: the systematic relationship between heritable advantage and population change over time, stable under perturbations of the underlying molecular machinery. Those invariants were real and trackable before their substrate was identified. When genetics arrived, it grounded Darwin's projection — explaining the mechanism of heritable variation — but it did not replace it. Natural selection remains the organizing concept of evolutionary biology because it picks out genuine structure at its own level of description, structure that the molecular account preserves rather than replaces.

The representational structure and the substrate interpretation came apart not only logically but historically — across decades of productive science. Such discontinuity is not uncommon. It may be the normal condition of productive inquiry: the projection does its work before the causal story arrives.

The same pattern recurs in current biology. Work on gene concepts has argued that the relationship between molecular substrate and higher-level biological description is mediated by developmental and cellular context in ways that resist simple read-off (Moss 2003). The representational structure of inheritance does its explanatory work at a level that the molecular account preserves rather than replaces.

## 3.3 The Vertical–Horizontal Criterion

The vertical cases might seem to support a reassuring picture: science converges, successor projections embed predecessors, and the direction of travel is toward a single unified description from which everything else is recoverable as a limiting case. That picture is tempting, and the logic of vertical embedding is precisely what shows why it is wrong.

The distinction between vertical and horizontal cases can be stated cleanly. Let $P_1$ be a projection with equivalence classes $C_1$ and invariants $I_1$. Let $P_2$ be a candidate successor with classes $C_2$ and invariants $I_2$. $P_2$ is vertically related to $P_1$ when there exists a specified limiting transformation under which the equivalence classes of $P_1$ are recoverable from those of $P_2$, and the invariants $I_1$ are recovered as limits of the invariants $I_2$. Kepler's ellipses stand in this relation to Ptolemy's circles: under the limit of vanishing eccentricity, the richer partition of conic sections reduces to the coarser partition of circles, and the orbital invariants tracked by the earlier projection are recovered as limits of those tracked by the later one. Einstein's spacetime stands in this relation to Newton's: under the limit of velocities well below c, the relativistic partition reduces to the classical one, and the classical invariants are recovered.

$P_2$ is horizontally related to $P_1$ when no admissible transformation within the space of physically or explanatorily motivated mappings recovers the equivalence classes of $P_1$ from those of $P_2$ under a limit, and no such transformation recovers the invariants $I_1$ as limits of $I_2$. The qualification matters: the claim is methodological, not metaphysical. The relevant space of transformations is the space of mappings warranted by the physics, the mathematics, or the explanatory practice of the domain — coarse-grainings justified by scale separation, reparameterizations grounded in symmetry, limits motivated by characteristic scales. A transformation constructed ad hoc to force limit-recovery is not admissible.

$P_1$, in such a case, partitions the state space in a way that has no counterpart in the finer-grained partition of $P_2$ below. Its equivalence classes are constituted at their own level of description, and its invariants live only there. The claim is not that physics cannot, in principle, entail the higher-level regularities — questions of metaphysical supervenience are not the target. The claim is that the higher-level invariants are not tractably available without the projection that defines the classes they live in, and the relation between the two projections is not one of limit-recovery under any admissible transformation.

Vertical and horizontal are therefore not descriptive labels but structural relations. A given pair of projections is vertically related just in case the limit-recovery condition is met; horizontally related just in case it fails. Mixed cases — where some invariants at the higher level are recoverable as limits and others are not — are partially vertical, partially horizontal, and the framework accommodates this. Section 5.1 examines the species problem in these terms: the Evolutionary Species Concept may stand in a partially vertical relation to biological, phylogenetic, and ecological species concepts, grounding some of the invariants that the other concepts track while leaving others horizontally constituted at their own levels.

## 4. Horizontal Cases

The vertical cases might suggest the familiar reductionist picture: finer-grained descriptions grounding coarser ones, physics eventually absorbing everything. The horizontal cases show why that picture is wrong.

The criterion of the previous section states the distinction precisely: in vertical cases, a successor projection contains the coarser classes of its predecessor as limits of its own, and the predecessor's invariants are recovered as limits of the successor's. In horizontal cases, the higher-level projection partitions the state space in a way that the finer-grained description below cannot reproduce under any admissible limit. The coarser partition is not a limit of the finer one. The invariants it makes visible have no counterpart in the finer-grained description.

The common structure of horizontal cases is multiple realizability. A kind is multiply realizable when the same higher-level property is instantiated by physically heterogeneous underlying states that share no relevant physical properties. Putnam introduced multiple realizability into the philosophy of mind to argue that mental states cannot be identified with specific physical states (Putnam 1967). Fodor extended the argument to the special sciences: economic, psychological, and biological kinds are multiply realizable, and the laws governing them cannot be reduced to physical law because no physical predicate covers all their instances (Fodor 1974). Multiple realizability has since generated extensive debate about whether it blocks reduction and how to distinguish genuine from trivial cases (Kim 1992; Shapiro 2000).

The projection framework offers something prior to those debates: a structural account of why multiple realizability produces level-specific invariants in the first place. Multiple realizability is the signature of a well-chosen projection — one whose omission of physical variation is constitutively appropriate to the level at which the invariant lives. The physically heterogeneous instances are multiple realizers of the same higher-level kind because the projection has correctly identified their physical differences as irrelevant. Constitutive omission is also constitutive concept formation: "money," "traffic wave," and "universality class" are not names for physical phenomena that were always there waiting to be labeled. They are concepts brought into being by projections that have correctly tracked what the world keeps stable at their respective levels of description.

### 4.1 Gresham's Law

Thomas Gresham observed that when legal tender laws require two currencies to be accepted at the same nominal value but they differ in commodity value, the inferior currency tends to circulate while the superior one is hoarded. Bad money drives out good. The Law ranges over a physically heterogeneous class of objects — shells, metals, paper, digital transfers — that share no physical properties. What makes them all "money" is a functional role defined by the economic projection.

A standard objection holds that "money" is merely a disjunction of physical states — gold or paper or digital entries — and that Gresham's Law is therefore just a shorthand for many separate physical regularities. The projection framework shows why this objection fails. "Money" is not a disjunction in the world; it is a single equivalence class constituted by the economic projection. A clipped gold coin and a paper note belong to the same equivalence class, not because they share a physical property but because the projection omits their physical differences and retains only the functional property of being a medium of exchange that legal tender laws require others to accept. The invariant — the systematic tendency for the inferior currency to circulate while the superior one is hoarded — holds across this class precisely because the projection constitutes that class by omitting what is irrelevant to it. The multiplicity of physical realizers is not a problem for the Law; it is the mark of a well-chosen equivalence class.

Zoom in on individual physical states, and Gresham's Law becomes explanatorily inaccessible. The equivalence classes that define "money" do not exist in the physical description. This omission is not ignorance of what happens at the physical level. It is a structural consequence of the fact that the relevant invariant is only constituted at the level of the economic projection.

### 4.2 Traffic Flow

Individual vehicles on a highway obey classical mechanics and their drivers' decisions. The physical description of each vehicle is, in principle, complete. And yet traffic jams propagate as waves with properties that belong to no individual vehicle: characteristic formation speeds, dissolution conditions, and density thresholds at which flow transitions from free to congested (Sugiyama et al. 2008). These wave dynamics are invariants of the traffic flow projection — stable across wide variations in the physical constitution of the vehicles, the road surface, and the drivers — and they are not tractably accessible at the level of individual vehicles.

The wave dynamics are multiply realizable: the same jam propagation behavior is instantiated by physically different collections of vehicles as long as density and spacing fall within the relevant thresholds. The omission of those physical differences is constitutive: it is not recoverable as a limiting case of a lower-level description, because the wave dynamics are only constituted at the aggregate level. Traffic dynamics involve no meaning, no intentionality, no institutions — just cars, roads, and mechanical decisions. And yet the relevant invariants exist at the aggregate level and cannot be recovered from the vehicle level as simple refinements.

This case is significant because it removes the standard social-complexity objection. Horizontal irreducibility does not depend on the presence of meaning, intentionality, or institutional structure. It arises wherever a projection constitutes equivalence classes that do not exist at the lower level — and that can happen in purely physical domains.

### 4.3 Universality Classes and the Renormalization Group

The most philosophically decisive horizontal case comes from physics itself.

Near a continuous phase transition — water approaching its critical point, a ferromagnet losing its magnetization at the Curie temperature — physical systems exhibit critical exponents: numbers describing how quantities like correlation length and heat capacity scale as the system approaches the transition. Systems with entirely different microscopic constituents share identical critical exponents (Wilson 1971; Wilson and Kogut 1974). Water and uniaxial magnets, built from different atoms obeying different force laws, belong to the same universality class.

The renormalization group reveals this by iterating the partitioning operation — applying successive coarse-grainings to the system and asking what remains invariant across them. At each step, short-range degrees of freedom are integrated out, and one asks what effective description of the remaining long-range behavior results. Most microscopic differences converge to zero under this procedure: they are genuinely irrelevant to the behavior at the critical scale. What remains invariant at the fixed point is the universality class — the set of all microscopic systems whose suppressed differences leave the critical structure unchanged.

The critical exponents are not visible in the microscopic description. A complete Hamiltonian for water does not contain them. They emerge only under the projection defined by renormalization group operations. The universality class invariant is only defined at the fixed point of the renormalization group, and no finite lower-level description makes it tractably available. Batterman has analyzed this as an asymptotic explanation, arguing that the explanation of universality requires mathematical operations that cannot be recovered from finite lower-level descriptions (Batterman 2002). The projection framework provides the structural account of what those operations do: they identify the partitioning within which the critical exponents are invariant, by systematically suppressing what is irrelevant to the behavior at the fixed point.

The renormalization group is therefore not merely an illustration of the invariant-tracking criterion. It is its precise physical implementation in domains where scale separation permits such analysis. A legitimate projection is one whose omissions leave the patterns it reveals undisturbed across reparameterizations of the underlying system. The renormalization group operationalizes this exactly — and the invariants it identifies satisfy the diagnostic of Section 2 in an especially strong form. Universality class membership is stable under an unbounded sequence of coarse-grainings (the stringent form of transformation-stability), is cross-model robust across structurally distinct microscopic theories (different Hamiltonians, different atomic constituents, different force laws), and supports counterfactual predictions about systems not yet measured. If the diagnostic is met anywhere, it is met here.

If horizontal irreducibility and multiple realizability appear here, in fundamental physics, they are not artifacts of social complexity or biological organization. They are structural features of how projection works.

### 4.4 Explanation Requires Omission

The horizontal cases challenge a deep intuition: that more information is always better, that finer-grained descriptions are always more accurate, and that progress always moves toward greater detail. These cases show otherwise.

Both vertical and horizontal omission serve the same explanatory purpose, but through different mechanisms. In vertical cases, omission enables explanation at a tractable level while remaining in principle connected to finer-grained descriptions through limiting relations. In horizontal cases, omission is constitutive of the level at which the invariant lives: it is not a simplification awaiting refinement but the condition under which the invariant exists at all.

Explanation requires omission. A projection succeeds when the variation it omits is genuinely irrelevant — stable under the transformations and reparameterizations relevant to the invariant being tracked. The economic projection omits physical composition and retains monetary function; that omission constitutes the equivalence class within which Gresham's Law holds. The traffic projection omits individual vehicle behavior and retains only aggregate flow properties; this omission defines the level at which wave dynamics are tractably accessible. The renormalization group suppresses microscopic details while preserving dimensionality and symmetry; that suppression defines the universality class. Omission is not incidental to explanation. It is the mechanism.

## 5. Pluralism, Integration, and the Coin

The framework is now in place. Three questions opened the paper; three resolutions follow. The first two — why superseded theories keep working, and why representational systems cannot see their own limits — are already answered, in effect, by the argument so far. A superseded projection that tracked genuine invariants is embedded in its successor as a limiting case; that is what Section 3 established. A representational system cannot register its own omissions because the omissions are not recorded within it; that is what Section 2.1 established through the structural parallel with perception. These dissolutions do not need to be restated. What remains is the second puzzle — the persistence of multiple frameworks in mature sciences — and the opening puzzle about probability that the paper has not yet revisited. Both require a closer look.

### 5.1 Why Mature Sciences Sustain Multiple Frameworks

Different projections of the same domain reveal different genuine invariants. The biological species concept tracks gene flow boundaries. The phylogenetic species concept tracks genealogical relationships. The ecological species concept tracks adaptive zones. Each constitutes distinct equivalence classes, makes different invariants visible, and induces constitutive omissions that are the price of what it makes tractably accessible. As Dupré (1993) and Ereshefsky (1992) argue, each tracks genuine biological structure that the others omit. The projection framework explains why: distinct projections define distinct equivalence classes, and the invariants visible under one projection are not recoverable from another without the projection-specific restructuring that defines the relevant classes.

Mayden (1997) has argued for the ontological priority of the Evolutionary Species Concept, which defines a species as a lineage maintaining its identity relative to other lineages. If Mayden is right, the other concepts are not independent horizontal alternatives but downstream projections of the deeper invariant that the ESC defines — a partial vertical structure within what appears to be purely horizontal pluralism. In the vocabulary of Section 3.3, the ESC stands in a partial limit-recovery relation to the other concepts, grounding some of the invariants they track (those recoverable as limits of lineage identity) while leaving others horizontally constituted at their own levels. The framework provides the vocabulary for that claim; whether the ESC succeeds in the role is a domain-specific question that no general account of representational structure can settle.

The persistence of multiple frameworks is not a failure of convergence. It is what the framework predicts for any domain rich enough to support multiple legitimate projections simultaneously. The species problem is not a problem. It is evidence that biology has found multiple genuine invariants.

Cartwright (1999) has argued that scientific pluralism is grounded in the ontological disunity of nature itself — capacities operating in specific domains, with no universal laws governing the whole. The projection framework aligns with Cartwright on the practice of science, but takes a more careful stance regarding its grounds. Multiple projections are warranted when each tracks genuine invariants that the others cannot reach. Whether a deeper projection could, in principle, unify what currently appears disunified is a question the framework leaves open. The patchiness Cartwright identifies may reflect the limits of present inquiry rather than a permanent feature of the world.

A more specific challenge has been pressed against this kind of pluralism from within the integrative-pluralist tradition. Integrative pluralism holds that the proliferation of explanatory frameworks in complex sciences is not a transitional defect but a warranted feature of mature inquiry, and that such frameworks can be coordinated without being reduced to a single unified theory (Mitchell 2003, 2023). Integrative pluralism makes two demands on its own coordination: *systematicity*, the coherent unification of diverse explanations, and *depth*, the provision of both global and local understanding. The tension is that satisfying both demands seems to require the very unification that pluralism resists. Miłkowski (2024) has recently argued that this tension is an inherent instability of integrative pluralism as currently formulated.

The projection framework dissolves this tension by showing that systematicity and depth are not competing demands but structural features of different kinds of projective progress. Vertical embedding provides systematicity: a successor projection that tracks the same genuine invariants must contain its predecessor as a limiting case, producing exactly the coherent unification across levels that systematicity requires. Horizontal irreducibility provides depth: invariants constituted at a given level of description are fully real and fully tractable at that level, providing exactly the local understanding that depth requires without demanding reduction to a unified lower-level account. The tension Miłkowski identifies is real — but it is a consequence of failing to distinguish these two structural modes. Once the distinction is in place, the tension dissolves.

### 5.2 Probability Revisited

The opening puzzle can now be closed. Before the coin flip, the probability of heads is one-half. After the coin lands, that probability vanishes. Where did it go?

The 1/2 is not a property of the underlying physical trajectory. It is an invariant of a projection that erases trajectory-relevant variation — one that partitions the space of possible flips by outcome while omitting the physical details that determine which outcome occurs. The invariant is not epistemic. It does not collapse to certainty as any individual's knowledge of the trajectory improves. All admissible formulations of the problem — formulations within the class of frameworks that satisfy the invariance diagnostic — assign the same value to a fair coin flip. The inter-agent claim is not that every agent in fact assigns 1/2; agents can be biased, miscalibrated, or working with defective frameworks. The claim is that every admissible formulation does. This invariance across admissible formulations is what distinguishes the 1/2 from a subjective credence. A credence varies with its holder's information. An invariant under an admissible formulation does not.

The 1/2 is also representation-independent within the relevant class of formalisms. A frequentist, a Bayesian, and a propensity theorist, working in incompatible formal frameworks, assign the same value to a fair coin flip. The value survives translation across the formalisms because it tracks something about the outcome structure of the flip rather than something about how any particular formalism encodes that structure. Inter-agent stability and formalism-independence together are exactly the cross-model robustness condition of the invariance diagnostic: the same regularity is detectable from structurally distinct projective angles. The 1/2 passes that diagnostic. It is an invariant in the sense that the paper has been using the term.

Nor is it merely instrumental. The invariant supports reliable intervention and prediction within the projection — it is the basis on which casinos are built, and risk is priced. That is precisely what distinguishes invariant structure from calculational convenience. An artifact of the description would not support stable prediction across independent applications. The 1/2 does.

What changes after the flip is not the agent's knowledge but the relevant question. The fine-grained physical projection becomes applicable where it was not before. The coarse-grained probabilistic projection becomes inapplicable where it was previously applicable. The 1/2 does not disappear because it was false. It becomes inapplicable because the outcome renders it redundant at a level of description. Inapplicability is not the same as falsity. A projection does not become false when a finer-grained description becomes available. It becomes redundant — superseded at a level of description by a different projection that is now applicable. The 1/2 was the invariant of the coarse-grained projection. The outcome is the invariant of the fine-grained one. Both are real. They are real at different levels of description, and the transition between levels is what the coin landing consists in.

The competing interpretations of probability — frequentist, Bayesian, and propensity — are not rival answers to a single question but distinct projections of the same underlying phenomena, each constituting different equivalence classes and making different invariants visible. The frequentist projection partitions the probability space by long-run relative frequency across a reference class of trials; its invariant is the distribution — the stable statistical structure that underlies actuarial tables, casino margins, and population-level prediction. Individual events fall outside its equivalence classes by design: single-case probability is not a deficiency of frequentism but a constitutive omission, the price of making distributional invariants tractably accessible. The propensity projection reaches where frequentism cannot: it makes the objective single-case tendency visible — the disposition of this physical setup, on this occasion, to produce this outcome. The distribution is downstream of propensities but is not what the projection foregrounds. The Bayesian projection operates at a different level entirely: it makes the rational structure of inference under uncertainty visible — how evidence should bear on credences, how agents should act given incomplete information — and is neutral on what probability is in the world. Each projection omits what the others foreground. Each tracks genuine structure. The centuries-long debate about which interpretation is correct may therefore be the wrong question — not a sign of philosophical failure but evidence that probability is rich enough to support multiple legitimate projections simultaneously, each resisting reduction to the others for the same structural reason that Gresham's Law resists reduction to particle physics.

One further observation bears on the argument as a whole. Because the omissions of any projection are invisible from within, revision cannot come from achieving a view outside the projection. It comes from the projection's failures — predictions the world does not confirm, anomalies that accumulate, outputs inconsistent with subsequent experience. The caloric projection was not abandoned because chemists could see the molecular variation it omitted. It was abandoned because outcomes at the thermodynamic level — the behavior of heat in reversible work cycles — were inconsistent with what the caloric framework predicted at that same level. The omitted variation remained invisible; the failure was entirely visible within the projection's own descriptive space. And anomalies alone never force abandonment of a working projection: the precession of Mercury's perihelion (Le Verrier 1859) remained unaccounted for by Newtonian mechanics for decades, without the framework being abandoned. An imperfect projection with known limitations is always preferable to no projection at all. The decisive event is not the recognition of the anomaly; it is the construction of a new projection that resolves it.

## 6. Conclusion

There is no view from nowhere. Every description partitions its domain into equivalence classes under constraints imposed by genuine invariant structure, omits variation within them, and tracks some invariants rather than others. The physicist omits the monetary function of coins to study their mass and charge. The economist omits their mass and charge to study their monetary function. Both are right. Neither is complete. Projection does not create invariants. It makes them tractably accessible by suppressing variation that is genuinely irrelevant to their stability. A description that omitted nothing would not be a description. It would be a reproduction.

The framework's deepest claim concerns the relationship between pluralism and realism. The standard picture treats these as in tension: if science is converging on a single correct description, pluralism is a transitional condition awaiting resolution. If pluralism is permanent, realism seems threatened. The projection framework dissolves this tension by inverting the standard picture. Perspectival structure is not a concession to complexity or a limitation of inquiry. It is the condition for invariant detection. Multiple projections are not evidence that the world resists a single description. They are evidence that the world is rich enough to support multiple genuine invariants simultaneously — invariants whose claim on reality is certified by their survival under the filtering conditions the diagnostic of Section 2 specifies, and which cannot all be made visible from a single projective stance.

A world rich in invariants cannot be exhausted by a single projection. Not pluralism despite realism. Pluralism because of realism.

The vertical-horizontal distinction is structural, not taxonomic. Vertical progress occurs when a successor projection embeds its predecessor as a limiting case: Keplerian ellipses contain Ptolemaic circles at eccentricity zero; population genetics contains the Darwinian selection projection as its upper level. Horizontal progress occurs when a projection reveals invariants specific to its level of description — invariants not recoverable from below: Gresham's Law, traffic wave dynamics, and universality classes are horizontal cases. The species problem illustrates that the vertical-horizontal distinction is not always clean: what appears to be a purely horizontal pluralism may have a partial vertical structure if one projection grounds the invariants that the others track.

The coexistence of multiple frameworks is not pathological but predicted: each projection satisfying the invariance diagnostic, each revealing genuine invariants the others omit, each inducing constitutive omissions that are the price of what it makes tractably accessible. The diagnostic does not merely separate successful from unsuccessful descriptions. It differentiates modes of failure according to which structural constraint is violated, thereby functioning not only as an interpretive framework but also as a constraint on admissible representations. The opening puzzles — the coin flip, the persistence of Newton, the proliferation of species concepts — all resolve here. They are instances of the same structure: representational enterprises that must omit variation to function, and whose omissions, when principled and constrained by genuine invariant structure, constitute the concepts through which the world becomes tractably accessible.

The framework does not tell scientists which projection to adopt for a given purpose. That is a domain-specific question that no general account of representational structure can settle in advance. What the framework provides is something prior: a precise account of what kind of question that is, what a principled answer must look like, and why the shape of scientific knowledge — its persistence across theory change, its productive pluralism, its characteristic blindness to its own limits — is not a puzzle to be explained away but a consequence of what representation, at any level, must do.